\documentclass{PoS}
\usepackage{multirow,url}
\let\olditemize\itemize
\renewcommand{\itemize}{
  \olditemize
  \setlength{\itemsep}{1pt}
  \setlength{\parskip}{0pt}
  \setlength{\parsep}{0pt}
}

\title{Constraining the Dark Matter decay lifetime with very deep observations of the Perseus cluster with the  MAGIC telescopes}
\ShortTitle{MAGIC Decaying Dark Matter Limits in Perseus}

\author{\speaker{J. Palacio}\\
        Institut de F\'isica d'Altes Energies (E-08193 Barcelona, Spain)\\
        E-mail: \email{jpalacio@ifae.es}}

\author{M.~Doro$^{a,b}$, M.~Vazquez Acosta$^{c,d}$, P.~Colin$^a$,
  C.~Maggio$^{b,e}$, J.~Rico$^f$, F.~Zandanel$^g$ on behalf of the
  MAGIC Collaboration\\
        $^a$ Max Planck Institute for Physics (Munich, Germany), \\
	$^b$ INFN-Padova (Padova, Italy), \\ 	
	$^c$ Instituto de Astrof\'isica de Canarias  (E-38205 La Laguna, Tenerife, Spain), \\
	$^d$ Universidad de La Laguna, Dpto. Astrof\'isica (E-38206 La Laguna, Tenerife,  Spain), \\
	$^e$ University of Padova  (Padova, Italy),\\
	$^f$ Institut de Fisica d'Altes Energies, (E-08193 Barcelona, Spain), \\
	$^g$ GRAPPA Institute, University of Amsterdam, (1098 XH Amsterdam, The Netherlands)
}

\abstract{We present preliminary results on Dark Matter searches from 
  observations of the
  Perseus galaxy cluster with the MAGIC Telescopes. MAGIC
  is a system of two Imaging Atmospheric Cherenkov Telescopes located
  in the Canary island of La Palma, Spain. Galaxy clusters are the
  largest known gravitationally bound structures in the Universe, with
  masses of $\sim10^{15}$ ~M$_\odot$. There is strong evidence that galaxy
  clusters are Dark Matter dominated objects, and therefore promising
  targets for Dark Matter searches, particularly for decay signals. 
  MAGIC has taken almost 300 hours of data on the Perseus Cluster between 2009 and 2015,
  the deepest observational campaign on any galaxy cluster performed so far 
  in the very high energy range of the electromagnetic spectrum. 
  We analyze here a small sample of this data 
  and search for signs of dark matter in the mass range between 100 GeV and 20
  TeV. We apply a likelihood analysis optimized for the spectral and
  morphological features expected in the dark matter 
  decay signals. 
  This is the first time that a dedicated Dark Matter optimization is applied 
  in a MAGIC analysis, taking into account the inferred Dark Matter distribution  
  of the source. The results with the full dataset analysis will be published 
  soon by the MAGIC Collaboration. }

\FullConference{The 34th International Cosmic Ray Conference,\\
		30 July- 6 August, 2015\\
		The Hague, The Netherlands}

\begin{document}

\section{Introduction}
Dark Matter (DM) is the most abundant component of  matter in
the Universe according to the 
current concordance cosmological model dubbed $\Lambda$CDM, making
up to 85\% of the total matter content~\cite{Bertone:2004pz}.
DM can be composed of one or more kind of particles, but as of now we have
only precisely determined the expected mass content through gravitational
interaction, at many different scales, from the galactic to the
cosmological one. The real nature of DM is unknown, included the mass range
where we expect DM to be.  
One of the main questions is whether DM is totally secluded in a dark sector
or it has some interaction with Standard Model particles. The Super-Symmetrical 
extension of the Standard Model of Particle Physics, for instance, gives a 
possible candidate, which is within the reach of the current experiments.
Within this framework, a natural DM candidate arises as the lightest
neutralino. This can be protected by a symmetry, and therefore stable and
self-annihilating, or alternatively, very long-lived possibly providing
decay signatures~\cite{Feng:2010gw}.

In this analysis, we investigate the possible signatures from DM
\emph{decays} by looking at a region in the sky where a very large density of DM
is expected: the core of clusters of galaxies~\cite{Voit:2004ah}.
In this work, we focus on  
the Perseus galaxy cluster, observed in Very High Energy gamma rays with
the MAGIC telescope for about 300~h in the years 2009-2015. 
Perseus is a cool-core cluster, the brightest in X-rays, and it is expected to
contain about $10^{14}$~M$_\odot$ in form of DM. 

The DM density distribution is usually parametrized following
the Navarro-Frenk-White (NFW, \cite{Navarro:1995iw}) profile, which reads as:
\begin{equation}\label{eq:NFWprofile}
  \rho(r)=\frac{\rho_s}{\frac{r}{r_s}\left(1+\frac{r}{r_s}\right)^2}
\end{equation}
where $\rho(r)$ is the DM density as a function of the distance $r$ from
the DM barycentre, and $r_{s}$ and $\rho_{s}$ are the characteristic
scale radius and density of the source. In the case of Perseus we have taken 
$(r_s,\rho_{s})=(0.477~\mbox{Mpc},7.25\times10^{14}\mbox{M$_\odot$Mpc$^{-3}$})$
following Ref.~\cite{SanchezConde:2011ap}. From the DM profile, 
the gamma-ray photon yield for decaying DM is computed using Eq.~\ref{eq:flux}:
\begin{equation}\label{eq:flux}
\Phi(E,\Delta\Omega)=\frac{1}{4\pi}\frac{1}{m_\chi\tau}\frac{dN_\gamma}{dE}\cdot
J(\Delta\Omega) \quad where \quad J(\Delta\Omega)=\int_{\Delta\Omega} d\omega \int_{los} \rho(r(s)) ds,
\end{equation}
where $m_\chi$ and $\tau$ are the DM particle mass and lifetime,
$dN_\gamma/dE$ is the number of photons globally produced as a
result of the DM decay (e.g. through pion decay after hadronization of
the quarks in the decay products) and $J(\Delta\Omega)$ is the
so-called astrophysical factor in the solid angle $\Delta\Omega$, which
is computed from the DM density profile as the integral over the line
of sight and the solid angle of the matter density of
Eq.~\ref{eq:NFWprofile}.

\begin{figure}
  \centering
  \includegraphics[width=0.9\linewidth]{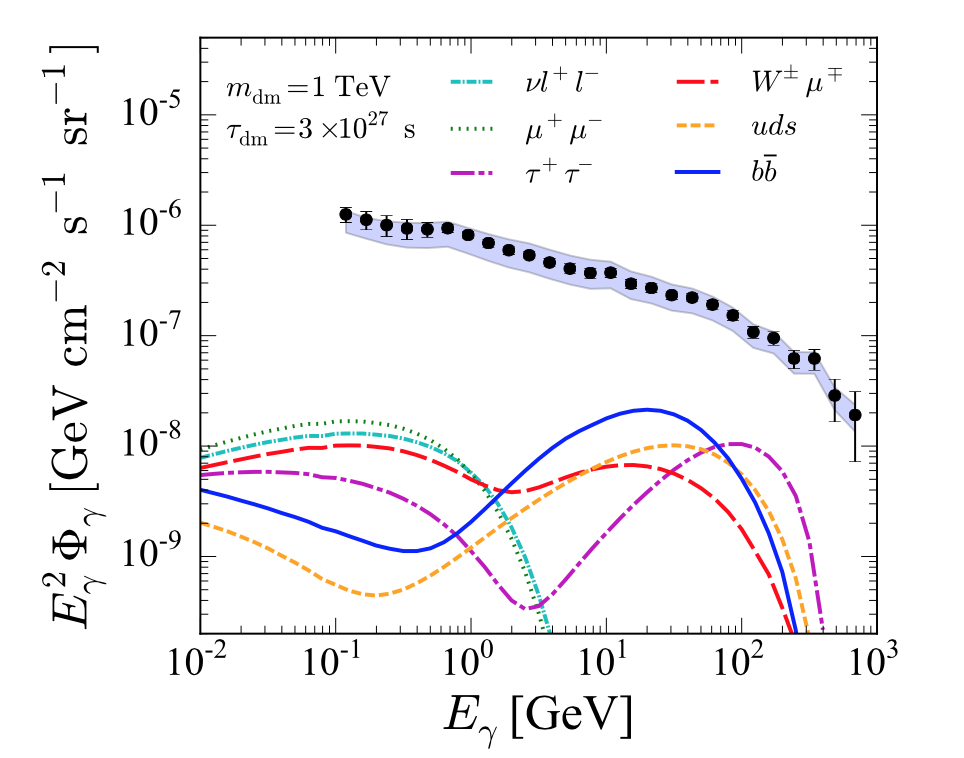}
  \caption{\label{fig:dmdecay} Gamma-ray spectra from different DM
    decay channel.  To
    provide a photon yield the mass of DM is fixed at 1 TeV and the
    decay lifetime to $3\times10^{27}$~s. Several decay channels are
    shown: (a) $\nu e\mu^-\mu^+$ $(\nu^- e\mu^-\mu^+)$
    and $\nu \mu e^- \mu^+$ $(\nu^- \mu^- e \mu^-)$,
    (b) $\mu^+\mu^-$,
    (c) $\tau^+\tau^-$,
    (d) $W^\pm \mu^\mp$ ,
    (e) $uds$ ($\bar{u}\bar{d}\bar{s})$, and
    (f) $b\bar{b}$.
    Data points with error bar and a band of the EGRB
    observed by Fermi-LAT is also shown. \underline{Figure is taken from Ref.~\cite{Ando:2015qda}}.}
\end{figure}
The result of decaying DM in which we are interested here
is the prompt production of gamma rays.
The gamma-ray yield depends largely on the primary DM nature, both in terms 
of flux
and spectral features. Fig.~\ref{fig:dmdecay} (taken from
Ref.~\cite{Ando:2015qda}) clearly shows the features of the
gamma-ray spectra for different DM decay scenarios. We also follow
Ref.~\cite{Ando:2015qda} here, to briefly introduce the decaying
DM phenomenology. There are three classes of decaying DM modes according
to the main products 
of the decay: $(i)$ mainly leptonic, $(ii)$ hadron and lepton, and
$(iii)$ mainly hadronic. 
Decaying DM models are also found in SUSY scenarios, if a mild violation of
the R-parity (RPV = R-Parity Violation) is allowed, such that the Lightest
Supersymmetric Particle (LSP) is not fully protected by the symmetry and
may decay to Standard Model particles.  There are several possibilities
to actually do the RPV, and according to the details, the LSP can be
a wino, a sneutrino, a gravitino or an axino. 
In all these cases, very energetic gamma rays would be generated
during the decay process.

\section{The Perseus Cluster}
Perseus is a Galaxy Cluster located at 77.7~Mpc (z=0.0183)
distance. With the above NFW parametrization, Eq.~\ref{eq:NFWprofile}, in case of 
decaying DM, 90$\%$ of the expected signal comes from a region of 1~deg, 
around the center of the cluster, which is 10 times larger
than the telescope point spread function ($\sim0.1^{\circ}$). 
For this reason, the signal-search region , i.e. $ \Delta\Omega$, 
must be extended and the exact aperture $\theta$ can be optimized considering that,
under larger solid angles, the irreducible background from primary
cosmic rays will also increase. We have therefore optimized
the signal region through a quality-factor (Q-factor) defined as:
\begin{eqnarray}
\label{eq:Qfactor}
Q\left(\Delta\Omega \right) & = & \frac{N_\gamma}{\sqrt(N_{bkg})} \propto \frac{\int los(\theta,\phi)\epsilon_{\gamma}(\theta,\phi)\theta d\theta d\phi}{\sqrt{\int\epsilon_{bkg}(\theta,\phi)\theta d\theta d\phi}}, \\
\nonumber & & \mathrm{where} \ \ los(\theta,\phi) = \int_{los} \rho(r(s)) ds.
\end{eqnarray}
The numerator in Eq.~\ref{eq:Qfactor} comes directly from
$J(\Delta\Omega)$ in Eq.~\ref{eq:flux} convoluted with
$\epsilon_{\gamma}$, i.e. the efficiency of the camera for gamma-like
events. The denominator is computed assuming a uniform flat background
rate through all the camera, again taking into account the efficiency,
$\epsilon_{bkg}$, for background events. We assumed both efficiencies
to be radially dependent, with respect to the center 
of the camera.
As a result, the highest Q-factor is 
obtained for $\theta=0.35^\circ$. The total
J-factor within this signal region is 
$J=5.6\times10^{18}$GeV cm$^{-2}$.
\begin{figure}[h!t]
  \centering
	\includegraphics[width=0.7\linewidth]{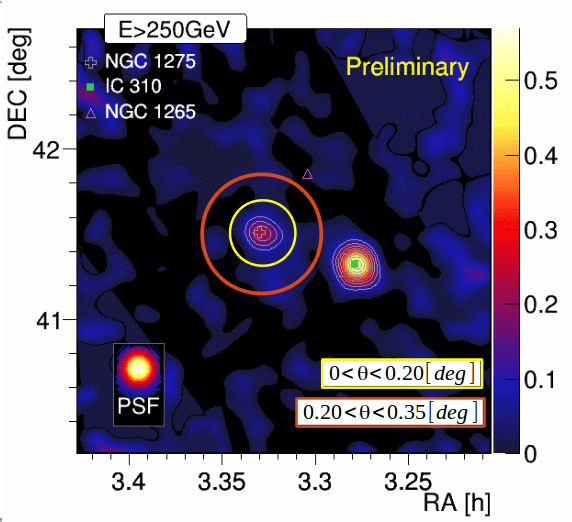}

\caption{\label{fig:wobble} 
		MAGIC sky map produced for the Perseus Cluster field of view. Some 
		significance can be seen for two member galaxies of the cluster,
		NGC1275 and IC310. On top of that, we plot the two signal regions
		defined for the analysis (see text).} 
\end{figure}
The tracking mode used in these observations, called wobble
mode~\cite{Fomin:1994aj}, allows us to obtain the background control
region simultaneously with the signal region. 

For each pointing direction, signal and background
are defined so that the regions where they are estimated have the same
size and are at the same 
distance with respect to the center of the camera.
In our case, the background events are 
obtained from a circular region of $0.35^\circ$ extension 
centered at a distance of $0.8^\circ$ away from the signal region.  
However, because of
the large extension of the DM expected signal, there is a ``leakage''
of the DM expected signal in the background control region, albeit small, which
we quantified as a few percent of the total expected signal 
and that is taken into account in our analysis.

\section{MAGIC observation and data reconstruction}
MAGIC\footnote{\url{http://magic.mpp.mpg.de}} is a system of two 17~m
diameter tessellated dishes working in stereoscopic mode to observe the
Cherenkov light coming from electromagnetic 
atmospheric showers initiated by VHE gamma rays
impinging the Earth. MAGIC is currently the instrument reaching the
lowest energy threshold among ground based 
gamma rays detectors~\cite{Aleksic:2014poa,Aleksic:2014lkm}. It is
located in the Canary island of La Palma (Spain) and is taking data since
2004. Its core science is focused on the study of the cosmic ray
origin in either galactic or extragalactic targets, but it is
well-known that cosmic gamma rays constitute also a probe for several
fundamental physics quests, including dark matter~(see, e.g.~\cite{Doro:2012xx}).
The Perseus campaigns proved very fruitful, providing 
the strongest limits on CR acceleration in the core of the cluster and the
cosmic ray to thermal pressure~\cite{Aleksic:2009ir,
  Aleksic:2011cp, Colin:2015}, the AGN
NGC1275 at the center of the cluster was clearly detected and
modeled~\cite{Aleksic:2011eb, Aleksic:2013kaa}, as well as interesting results came from the peculiar
radio-galaxy IC310 located at 0.6~deg from the Perseus center,
bringing an interesting clue in the mechanism of acceleration of cosmic
rays close to black holes~\cite{Aleksic:2010xk, Aleksic:2013bya, Aleksic:2014xsg}. Here, we focus on the
signatures from DM decays coming from the cluster.

During the 6-year observational campaigns, MAGIC has gone through several hardware
upgrades, and therefore different data samples are analyzed using 
different instrumental response function (IRFs, mainly effective area and
energy migration tables), the details of which will be reported in the
incoming publication of the MAGIC collaboration, under
preparation. For each data sample, pixel-wise data are cleaned and 
combined into global event parameters expressed with the Hillas
formalism~\cite{Hillas:1985}, and assigned a parameter called ``hadronness'' built
through a classification method called Random Forest (RF), through
comparison with a tuned Monte Carlo sample. The RF also assigns the
events estimated energy and direction values. In the further
analysis steps, data is analyzed through the full likelihood method
described in~\cite{ALEKSIC:2014foa}. The input of the (binned) likelihood are: the list of
the energy of photons coming from the signal region and the background
control region (after various quality cuts, specially the hadronness
one), as well as the IRFs above mentioned for that period. The
likelihood is built defining the null hypothesis as the 
case where no excess over the background is found, while the test
hypothesis considers the flux computed using Eq.~\ref{eq:flux} \emph{and}
under the hypotheses of two different pure DM decays: one into
$b\bar{b}$ and one into $\tau^+\tau^-$ often used in the literature as
representative of soft and hard spectra respectively. Other channels
have a spectral distribution normally in between these two. The
analytical formulas for the photon yield in the two cases are taken
from~\cite{Cembranos:2010dm}.

To build the correct IRFs for each subsample, we developed a new tool 
to selectively choose from a diffuse sample of Monte
Carlo (globally covering the 3 deg of the camera FoV) events, those
that match the
expected distribution of signal events in the data. This procedure
will also be described in the incoming MAGIC publication. 

Summarizing, we have prepared
$7\left(\mathrm{Period}\right)\times2\left(\mathrm{Zd}\right)\times2\left(Ring\right)=28$ different subsamples, 
each of them with different IRFs,
which are later on combined together using the Full Likelihood
formalism. In the above list, "Period" are the stable periods 
of the Telescopes, 
the two bins in "zenith angle (Zd)" are
appropriate because both the effective area and energy resolution are
affected
and "Ring" stands for the division of our signal region into two
concentring rings.
The reason for this last spatial binning is first, the dependence 
of the camera response at large distances with respect to its center, 
and most important, 
NGC1275's expected emission, lying at the center of the cluster. 
The size of the first bin was chosen so that more than 95$\%$ of 
NGC1275's emission is expected to be in the inner ring~\cite{Aleksic:2014lkm}.
This astrophysical contamination needs a special treatment 
in the likelihood analysis and details on this will be 
reported in the upcoming full data analysis publication.

\section{Results}
We present the results on the decay life time constraints obtained 
with a small subsample of the Perseus Cluster data taken with MAGIC.
We analyze data taken during the period from 2013.07.27
to 2014.08.05, with zenith angles between 5$^\circ$ to 50$^\circ$ 
and only one pointing mode (we have two pointing modes).
This sums up to 
$1\left(\mathrm{Period}\right)\times2\left(\mathrm{Zd}\right)\times1\left(\mathrm{Ring}\right)=2$
subsamples with different IRFs. 
We have made all the consistency checks for the 
robustness of the reconstruction, which will be later on 
applied to the complete data set analysis.

We found no excess in this region of the sky.
The total J-factor in the case of
decaying DM corresponds to  about $J(0.2<\theta<0.35) =
1.8\times10^{18}$GeV cm$^{-2}$. By fixing a DM mass
and a decay channel, we derive, from our null detection, 
a lower limit on the DM decay lifetime using
eq.~\ref{eq:flux}. For the decay channel, we used the analytical computation
of~\cite{Cembranos:2010dm} (based on PITHYA), for two pure decay
channels: $b\bar{b}, \tau^+\tau^-$.  
We repeat the procedure for several DM masses in the range 
160 GeV to 200 TeV and compute again the lower limits.

Our results, obtained with a subset of 12.40~h,
provide a lower limit on the DM decay lifetime of $10^{24}$~sec at 100
GeV and $10^{26}$~sec at 1 TeV. With the full data sample analysis
we expect an improvement in sensitivity of a factor 4 with compared
to the present results.
\section{Summary and Conclusions}
We report here the lower limits on the dark matter decay lifetime obtained
with a subset of data taken with the MAGIC telescope of the Perseus galaxy cluster.
The full dataset comprises
300~h in 6 years with MAGIC and will be published soon.
A binned likelihood method will be used to
combine results taken under different experimental conditions,
including different data taken periods, different
observation modes, different zenith angles and radial distances from
the center of the source. 

\subsection*{Acknowledgement}
\footnotesize{We would like to thank the Instituto de Astrof\'{\i}sica
  de Canarias for the excellent working conditions at the Observatorio
  del Roque de los Muchachos in La Palma. The financial support of the
  German BMBF and MPG, the Italian INFN and INAF, the Swiss National
  Fund SNF, the ERDF under the Spanish MINECO (FPA2012-39502), and the
  Japanese JSPS and MEXT is gratefully acknowledged. This work was
  also supported by the Centro de Excelencia Severo Ochoa
  SEV-2012-0234, CPAN CSD2007-00042, and MultiDark CSD2009-00064
  projects of the Spanish Consolider-Ingenio 2010 programme, by grant
  268740 of the Academy of Finland, by the Croatian Science Foundation
  (HrZZ) Project 09/176 and the University of Rijeka Project
  13.12.1.3.02, by the DFG Collaborative Research Centers SFB823/C4
  and SFB876/C3, and by the Polish MNiSzW grant 745/N-HESS-MAGIC/2010/0.}

\end{document}